\documentclass[fleqn,usenatbib]{mnras}

\usepackage{newtxtext,newtxmath}

\usepackage[T1]{fontenc}
\usepackage{url}
\usepackage{hyperref}
\usepackage{xcolor}
\usepackage{afterpage}

\usepackage{lineno}

\usepackage{bm}

\newcommand{\jade}{\texttt{JADE}}
\newcommand{\sbilens}{\texttt{sbi\_lens}}
\newcommand{\wcdm}{\textit{w}CDM}

\DeclareRobustCommand{\VAN}[3]{#2}
\let\VANthebibliography\thebibliography
\def\thebibliography{\DeclareRobustCommand{\VAN}[3]{##3}\VANthebibliography}

\usepackage{graphicx}	
\usepackage{amsmath}	

\title[Joint convergence map and cosmology inference]{Joint inference of weak lensing convergence map and cosmology with diffusion models}
\author[B. Remy et al.]{
Benjamin Remy,$^{1,2,}$\thanks{E-mail: bremy@uchicago.edu}
Chihway Chang,$^{1,2,3}$
Rebecca Willett$^{1,4,5}$
\\
$^{1}$NSF-Simons AI Institute for the Sky (SkAI), 172 E. Chestnut St., Chicago, IL 60611, USA\\
$^{2}$Department of Astronomy and Astrophysics, University of Chicago, Chicago, IL 60637, USA\\
$^{3}$Kavli Institute for Cosmological Physics, University of Chicago, Chicago, IL 60637, USA\\
$^{4}$Department of Statistics, University
of Chicago, USA\\
$^{5}$Department of Computer Science, University of Chicago, USA\\
}

\date{Accepted XXX. Received YYY; in original form ZZZ}

\pubyear{2026}

\begin{document}
\label{firstpage}
\pagerange{\pageref{firstpage}--\pageref{lastpage}}
\maketitle

\begin{abstract}
We present a method for joint inference of cosmological parameters and convergence maps from weak lensing observations, targeting the full posterior conditioned on the observed shear field. 
Our approach uses implicit inference 
with diffusion models, learning the joint distribution from simulations, without the need to have an explicit and differentiable forward model for gradient-based MCMC sampling. We introduce a transformer-based architecture that operates in pixel space and treats cosmological parameters as additional tokens in a unified sequence, enabling efficient multimodal processing within a single network. At inference time, the trained model generates posterior samples of joint convergence maps and cosmological parameters conditioned on observed noisy shear fields.
We demonstrate the method on simulated weak lensing data generated from log-normal fields in a \wcdm\ cosmology. The model accurately reconstructs convergence maps and recovers cosmological posteriors that agree with traditional MCMC, while remaining well calibrated across the prior, with a MIRA calibration score of $0.635 \pm 0.017$ on the joint posterior (where $0.667$ is optimal). The inferred fields reproduce the correct two-point statistics as well as non-Gaussian statistics such as the one-point distribution. This work establishes diffusion-based implicit inference as a viable route toward full field-level cosmological analyses, paving the way for applications to more realistic, non-differentiable simulators.
\end{abstract}





\section{Introduction}
Ongoing and upcoming Stage-IV galaxy surveys, including DESI \citep{aghamousa2016desi}, Euclid \citep{laureijs2011euclid}, the Vera C. Rubin Observatory's Legacy Survey of Space and Time \citep[LSST,][]{ivezic2019lsst} and the Roman Space Telescope's High-Latitude Wide-Area Survey \citep[HLWAS,][]{spergel2015wide} will map the large-scale structure of the Universe with unprecedented precision. Harnessing the full potential of these datasets requires analysis methods capable of exploiting all available information, particularly on small, non-linear scales where a vast number of observable modes carry significant constraining power on cosmological parameters.

Weak gravitational lensing, the subtle coherent distortions of galaxy shapes caused by the gravitational potential of the large-scale structure, is one of the most powerful probes of cosmology \citep{Mandelbaum2018}. Traditional cosmological analyses rely on summary statistics, primarily the two-point correlation function of galaxy shapes, to extract cosmological information from lensing surveys. While highly successful, e.g. analyses from DECADE+DES
Y3 \citep{anbajagane2025dark}, KiDS-Legacy \citep{wright2025kids}, HSC Y3 \citep{li2023hyper}, these approaches are known to discard information contained in the non-Gaussian features of the matter distribution, particularly on small scales where nonlinear structure formation generates rich higher-order correlations.

To go beyond the two-point function and access the non-Gaussian information present in weak lensing data, a variety of higher-order statistics have been developed. These include peak counts, wavelet and scattering transforms, Minkowski functionals, moments, and three-point statistics \citep[see e.g.][and references therein]{ajani2020constraining,cheng2020new,liu2019constraining,kratochvil2012probing,takada2004cosmological}. These approaches have been shown to tighten cosmological constraints relative to two-point analyses alone. However, any fixed set of summary statistics inevitably leaves some fraction of the available information unexploited. Full-field approaches, which operate directly on the observed maps rather than compressing them into predefined summaries, offer a principled way to capture all statistical information simultaneously.

One line of work uses neural networks to perform full-field cosmological inference directly from weak lensing maps, learning near sufficient compressed summaries of the field from which the cosmological posterior is then sampled  \citep{jeffrey2021likelihood, Lanzieri_2025, jeffrey2025dark, thomsen2025dark}.
These methods can in principle capture all statistical information present in the observations, including non-Gaussian signatures, thereby reducing uncertainties on cosmological parameters beyond what analytical summary statistics alone can achieve. However, they marginalize over the underlying convergence field rather than reconstructing it, so the map itself is not recovered as part of the inference. 

A separate line of work, also based on neural networks, instead targets field reconstruction in isolation. mass mapping methods \citep{remy2023probabilistic} aim to recover the convergence field from noisy shear measurements, and analogous approaches have been developed for reconstructing initial conditions from late-time tracers \citep{legin2024posterior, Doeser_2025}. These methods treat the inverse problem on its own, with cosmological parameters held fixed rather than jointly inferred. Most existing neural network based methods thus address either the reconstruction or the inference problem while holding the other fixed, or resort to marginalization.

In this paper, we introduce \texttt{JADE}\footnote{\url{https://github.com/b-remy/jade}} (Joint Architecture for fielD and cosmological parameter Estimation), an alternative approach based on implicit inference, in which we learn a conditional diffusion model that directly samples from the posterior distribution of physical fields and cosmological parameters, conditioned on the observed data. Rather than requiring a differentiable forward model and MCMC sampling, our method learns this conditional distribution directly from simulations, allowing for arbitrarily complex cosmological simulators. This enables both mass map reconstruction and cosmological parameter inference in a single, unified framework. Related to this work, \cite{cuesta2024joint} developed a joint inference pipeline combining two separate generative models, one for inferring the initial density field, and the other for the cosmological parameters. While we could use this approach to sample from our target joint distribution between the convergence and cosmology, we propose a method that learns the joint posterior with a single conditional diffusion model. We demonstrate that this implicit approach can recover accurate posterior distributions over both the convergence field and cosmological parameters, opening the door to full-field analyses with realistic, non-differentiable simulators.

The paper is organized as follow: in Section~\ref{sec:weak-lensing} we recall how weak lensing serve as a cosmological probe and how we mocked convergence fields with log-normal simulations, in Section~\ref{sec:method} we present our diffusion model framework and our transformer architecture for joint inference of cosmology and mass maps, and in Section~\ref{sec:dataset} we describe the simulation setting we used to train our diffusion model, and we present the results of joint inference in Section~\ref{sec:results}. We conclude in Section~\ref{sec:discussion}.

\section{Weak gravitational lensing}\label{sec:weak-lensing}

\subsection{Primer on convergence and shear}

The large-scale structures of the matter distribution deflect the paths of photons emitted by distant galaxies due to their gravitational potential, inducing
subtle but coherent distortions in their observed shapes. With this effect being faint, we call it the weak regime of gravitational lensing, or weak lensing. These distortions are described by two quantities: the convergence $\kappa$, which corresponds to an isotropic magnification of galaxy images, and the shear $\gamma=\gamma_1 + i \gamma_2$, which produces anisotropic stretching.

The convergence is directly related to the projected matter density, or mass, along the line of sight. As such, a map of the convergence field is often referred to as a \textit{mass map}. For a population of source galaxies distributed in redshift according to a density $n(z)$, $\kappa$ can be written as a weighted integral of the three-dimensional matter overdensity $\delta$ over the comoving distance $\chi$,

\begin{equation}\label{eq:convergence}
    \kappa(\bm{\vartheta}) = \frac{3H_0\Omega_m}{2c^2}\int^{\chi_\text{lim}}_0 d\chi\frac{q(\chi)}{a(\chi)}f_{K}(\chi)\delta(f_K(\chi)\bm{\vartheta}, \chi),
\end{equation}
where $\bm{\vartheta}$ is the angular coordinate on the sky, $H_0$ is the Hubble constant, $\Omega_m$ the matter density parameter, $a(\chi)$ the scale factor, $f_K$ the comoving angular diameter distance, and $q(\chi) = \int^{\inf}_\chi d\chi^\prime n(\chi^\prime) f_K(\chi^\prime - \chi)/f_K(\chi^\prime)$ is the lensing efficiency kernel. This expression also makes it explicit that the convergence field depends on cosmological parameters governing the geometry and growth of structure \citep{Kilbinger2015}.

While convergence encodes the projected mass, it is not directly observable. In practice, we measure shear, which can be estimated from the ellipticities of galaxies. 

Both the convergence and shear are obtained from the lensing potential $\psi$ as second derivatives. Adopting the flat-sky approximation, in which the survey patch is treated as a plane with Cartesian coordinates, these take the form
\begin{equation}
    \kappa = \frac{1}{2}\Delta\psi \; , \; \gamma_1 = \frac{1}{2}(\partial_1^2\psi - \partial_2^2\psi) \; , \; \gamma_2 = \partial_1\partial_2\psi,
\end{equation}
where $\Delta = \partial_1^2 + \partial_2^2$. In the same approximation, we can convert between them through the following relation in Fourier space \citep{Kaiser1993}
\begin{align}
    \tilde{\kappa}_E + i\tilde{\kappa}_B &= \left(\frac{k_1^2 - k_2^2}{k^2} + i\frac{2k_1 k_2}{k^2}\right) \left(\tilde{\gamma}_1 + i\tilde{\gamma}_2\right) \nonumber \\
    &= \mathbf{P} \left(\tilde{\gamma}_1 + i\tilde{\gamma}_2\right),
\end{align}
where $\kappa_E$ and $\kappa_B$ are the E- and B-modes of the convergence, and a tilde denotes the Fourier transform. The forward model relating the convergence to the observed noisy shear is therefore
\begin{equation}
    \gamma_\text{obs} = \mathbf{F}^{-1}\mathbf{P^\dagger F}\kappa + n,
\end{equation}
with $\mathbf{F}$ the Fourier operator,  $\mathbf{P}^\dagger$ the Hermitian conjugate of $\mathbf{P}$, and $n\sim\mathcal{N}(0,\Sigma_n)$ the shape noise. 

\subsection{Log-normal mass maps}\label{sec:log-normal}

To test our pipeline, we need to generate a large number of weak lensing convergence maps to train the diffusion model. There is a wide spectrum of methods that one can adopt to generate these maps ranging from N-body simulations, e.g. the Gower street simulations \citep{jeffrey2025dark} or the CosmoGrid simulations \citep{kacprzak2023cosmogridv1}, to simple Gaussian maps given a power spectrum \citep{xavier2016flask, tessore2023glass}. In this paper, we take an intermediate approach and simulate convergence maps with a 2-dimensional log-normal model \citep{coles1991lognormal}. 

The log-normal field is a phenomenological model for the convergence, which has been demonstrated to be able to mock the non-linear growth of structures \citep{Xavier_2016, clerkin}. While log-normal fields do not fully replicate the complexity of N-body simulations, particularly at small scales, they are very fast to generate and embed non-Gaussian information in the fields, making them very useful for demonstrating novel field-level inference methods. We refer the reader to \cite{Lanzieri_2025} and \cite{Zeghal_2025} for an in-depth description of the efficient generation of log-normal fields, and only outline its main steps here. 

First, a Gaussian random field $\kappa_\text{GRF}$ is generated with the expected convergence non-linear power spectrum, and with expected cross-correlations across redshift bins. Then this field is exponentiated and shifted with a parameter $\lambda$, generating a log-normal field $\kappa_\text{LN}$ such that
\begin{equation}
    \kappa_\text{LN} = e^{\kappa_\text{GRF}} - \lambda(z, \theta).
\end{equation}
Because the exponentiation changes the pixel correlations, the field needs to be normalized so that it matches the expected correlation function. Typically, we construct the correlation function of $\kappa_\text{GRF}$ such that its transformation into a log-normal field yields correlation functions that reproduce the theoretical predictions. The shift parameter $\lambda$ is determined from perturbation theory calculations, using the \texttt{CosMomentum} code \citep{friedrich2020primordial}, to ensure the log-normal field reproduces the correct skewness. The dependency on cosmology and redshift of the shift parameter induces information in the fields beyond the power spectrum, hence the presence of non-Gaussian information.

While applying our method to real data will ultimately require training on N-body simulations, the log-normal model is sufficient for the purposes of this work as it provides a fast, controllable forward model that produces fields with realistic non-Gaussian structure, allowing us to develop and test the joint inference framework before scaling to more expensive simulators.

\section{Methods} \label{sec:method}

In this section, we provide an overview of field-level inference and where our approach stands in Section~\ref{sec:field-level}. Next, we describe our diffusion model framework in Section~\ref{sec:diffusion-model} and our proposed neural network architecture in Section~\ref{sec:jade}. We finally explain how to condition the inference on observed fields in Section~\ref{sec:amortized-posterior}.

\subsection{Field-level inference}\label{sec:field-level}
In the context of weak lensing, field-level inference is formulated as a forward modeling problem: starting from cosmological parameters $\theta$ and initial condition over-densities $\delta_\text{IC}$, a physical model predicts the convergence field $\kappa$, from which the observed noisy shear field $\gamma_\text{obs}$ is generated. The goal is then to infer a posterior that jointly constrains the cosmology and one or more of these underlying fields.

A prominent example is the Bayesian Origin Reconstruction from Galaxies framework \citep[BORG,][]{jasche2013bayesian, lavaux2019systematicfreeinferencecosmicmatter}, which targets the joint posterior over cosmological parameters and initial conditions, $p(\theta, \delta_\text{IC} \mid \gamma_\text{obs})$, by sampling within a Markov Chain Monte Carlo (MCMC) framework. This formulation has shown promise for inferring cosmological parameters such as ($\Omega_m$, $\sigma_8$) \citep{porqueres2023fieldlevelinferencecosmicshear}, primordial non-Gaussianity \citep{chen2024probing, andrews2023bayesian}, and BAO parameters \citep{bayer2026field}. Beyond its role in constraining cosmology, reconstructing the initial conditions has independent scientific value: when coupled with a forward model, the recovered initial density field can be used to predict fields that are not directly observable, such as the late-time matter density field, the convergence map, or the detailed structure of the cosmic web, like  clusters, filaments, and voids, providing observational access to quantities that have traditionally been studied only through simulations. It also enables direct validation of the assumed cosmological model through posterior predictive checks. However, these approaches rely on \textit{explicit inference}: they require a differentiable forward model linking the initial conditions to observables, which limits the complexity of the physics that can be employed, and they sample the posterior using MCMC, which faces well-known challenges in the high-dimensional spaces characteristic of field-level inference. Gradient-based samplers such as Hamiltonian Monte Carlo \citep[HMC;][]{radford2021learningtransferablevisualmodels, betancourt2018conceptualintroductionhamiltonianmonte} can accelerate the sampling when a differentiable forward model is available, but do not lift the underlying restrictions on model complexity. \citet{omori2026practicalfieldlevelinferenceweak} shows that sampling 3D initial conditions from observing the 2D cosmic shear field is particularly challenging when using particle-mesh solvers, and is feasible only with careful preconditioning of the sampler.

The second strategy, known as \textit{implicit inference}, bypasses the need for an explicit likelihood and a differentiable forward model by using generative models trained on simulations to learn the relationship between observations and parameters directly from simulations.

A further distinction concerns what is being inferred. \textit{Marginal inference} targets the cosmological posterior $p(\theta \mid \gamma_\text{obs}) = \int p(\theta, \kappa \mid \gamma_\text{obs})\, \mathrm{d}\kappa$, integrating out the field and providing constraints on cosmological parameters only. \textit{Joint inference}, on the other hand, targets the full posterior $p(\theta, \kappa \mid \gamma_\text{obs})$, simultaneously constraining the cosmology and reconstructing the convergence field. Implicit inference methods have been explored for weak lensing \citep{Jeffrey_2020, Lanzieri_2025, jeffrey2025dark, thomsen2025dark}, but existing approaches all target the marginal posterior and do not provide an estimate of the convergence field.

Our method falls within the implicit inference framework, but unlike previous work, we target the full joint posterior $p(\theta, \kappa \mid \gamma_\text{obs})$, enabling both cosmological parameter inference and convergence field reconstruction within a single model. We choose to infer the convergence field as a first step to demonstrate our approach, as inferring jointly the initial conditions and cosmology $p(\theta, \delta_\text{IC} \mid \gamma_\text{obs})$ is more challenging, as one needs to infer a 3D field from a 2D observed field. Future work will extend our proposed architecture to the joint inference of initial conditions and cosmological parameters.

\subsection{Diffusion models} \label{sec:diffusion-model}

\begin{figure*}
    \centering
    \includegraphics[width=0.8\linewidth]{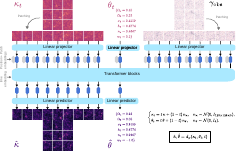}
    \caption{Joint field and cosmological denoising transformer architecture inputs a noisy convergence map, noisy cosmological parameters, and observed shear field, and outputs denoised field and cosmology (hat denotes estimated variable). Fields are first patched and then linearly projected into embedding vectors, and summed to their positional encoding. Time embedding modulates each transformer block.
}
    \label{fig:jade}
\end{figure*}

The first step of our pipeline concerns learning the joint distribution $p(\kappa, \theta)$ with a diffusion model.
For simplicity, we introduce a joint variable $x = [\theta, \kappa]$. Diffusion models \citep{ho2020denoisingdiffusionprobabilisticmodels, song2021scorebasedgenerativemodelingstochastic} and flow models \citep{lipman2023flowmatchinggenerativemodeling} are generative models that map a base distribution $p_0$, such as a multivariate Gaussian, to a target distribution $p_1$. Diffusion and flow models differ in their specific training and sampling procedure, but are fundamentally the same generative model framework as unified in \cite{albergo2023stochastic}, so we interchangeably use diffusion and flow in this paper.  Here, we adopt the flow matching framework where we introduce a noisy variable $x_t = tx_1 + (1-t)x_0$, which linearly interpolates between the base and the target distribution, and where the generation is performed by sampling from the base distribution $x_0 \sim p_0$ and solving the ordinary differential equation (ODE)

\begin{equation}\label{eq:ode}
    \dfrac{dx_t}{dt} = v_t(x_t),
\end{equation}
where $v_t$ is called the marginal velocity field and can be estimated with a neural network $v_\varphi(x_t, t)$, parameterized by weights $\varphi$. The marginal velocity field $v_t(x_t)$ is generally intractable, but flow matching provides a tractable way to learn it. The key idea is to introduce a conditional velocity field $v_t(x_t\mid x_0,x_1)$, i.e. conditioned on the initial and final state of the trajectory, that we get by differentiating the interpolant such as

\begin{equation}
    v_t(x_t\mid x_0, x_1) = \frac{dx_t}{dt} = x_1 - x_0.
\end{equation}

The flow matching objective is then a regression loss with respect to this velocity field, such that for a neural network $v_\varphi$ we need to minimize the mean squared error loss

\begin{align}\label{eq:loss-velocity}
    \mathcal{L}(\varphi) 
    & =\mathbb{E}_{p(t)p(x_1)p(x_0)} \left[ \| v_\varphi(x_t, t) - v_t(x_t\mid x_0, x_1) \|^2 \right] \nonumber \\
    &= \mathbb{E}_{p(t)p(x_1)p(x_0)} \left[ \| v_\varphi(x_t, t) - (x_1 - x_0) \|^2 \right],
\end{align}
where $t\sim p(t)$ is the time distribution during training. This loss is being minimized when $v_\varphi$ equals the conditional expectation $\mathbb{E}_{p(x_0,x_1\mid x_t)}\left[ v_t(x_t\mid x_0, x_1) \right] = v_t(x_t)$, which is exactly the marginal velocity field.
Alternative objectives have been proposed for this task \citep{ho2020denoisingdiffusionprobabilisticmodels, song2021scorebasedgenerativemodelingstochastic, albergo2023stochastic}, and in particular \cite{li2026basicsletdenoisinggenerative} showed that while the different objectives, including \autoref{eq:loss-velocity}, lead to equivalent performance,  parameterizing the neural network so that it predicts the denoised sample directly instead of the velocity can substantially improve the performance. To this end, we introduce a denoiser $d_\varphi(x_t, t)$, and parameterize the velocity field such that

\begin{equation}\label{eq:velocity-denoiser}
    v_\varphi(x_t, t) = \dfrac{d_\varphi(x_t, t) - x_t}{1 - t}.
\end{equation}

Note that this expression is not defined for $t=1$, so we clip the denominator with a strictly positive value following \cite{li2026basicsletdenoisinggenerative}. The loss function expressed in term of the denoiser can be found in \autoref{eq:jade-loss}. The motivation for the denoiser parameterization relates to the low-dimensional manifold structure of physical data. Convergence fields, like natural images, occupy a low-dimensional manifold within the ambient pixel space. Predicting a noised quantity like the velocity $v = x_1 - x_0$ requires the network to faithfully represent signals that span the full ambient space, placing heavy demands on network capacity, particularly for large patch sizes. Predicting the clean sample $x_1$ instead allows the network to focus on the manifold structure of the data, which is a considerably easier task. As demonstrated by \cite{li2026basicsletdenoisinggenerative}, this distinction goes beyond a simple reweighting of the loss: direct velocity or noise prediction can break down entirely in high-dimensional settings, whereas clean-sample prediction remains effective even with limited network width.

The framework described above is general as it defines how to learn and sample from a high-dimensional distribution using a neural network that parameterizes the velocity field. What remains to be specified is the architecture of this neural network, which must be able to operate jointly on both the convergence field $\kappa$ and the cosmological parameters $\theta$. We address this in the next section.

\subsection{A transformer architecture for field and parameters inference} \label{sec:jade}

As we aim to learn the joint distribution $p(\kappa, \theta)$, we need an architecture which takes both $\kappa$ and $\theta$ as input and produces both as output. The usual architecture used for diffusion models involving images or volumes is the U-Net \citep{ronneberger2015unetconvolutionalnetworksbiomedical} thanks to the efficiency of convolutional layers. While this architecture has been widely used for astrophysical application \citep{remy2023probabilistic, adam2022posterior, legin2024posterior}, it is not convenient to work with cosmological parameters as well. Transformer-based architectures \citep{vaswani2023attentionneed} however, initially developed for text, have recently demonstrated equivalent performances for images \citep{dosovitskiy2021imageworth16x16words} and multimodal data \citep{radford2021learningtransferablevisualmodels}. They have already been used for multimodal tasks in astrophysics \citep{parker2025aion1omnimodalfoundationmodel}. Transformer architectures along with diffusion models have also been used in the context of simulation-based inference to model joint and arbitrary conditional distributions in \citep{gloeckler2024all}.

We introduce \texttt{JADE}, a Joint Architecture for fielDs and cosmological parameter Estimation. It is a pixel-space vision transformer augmented with cosmological information. \cite{li2026basicsletdenoisinggenerative} recently demonstrated that vision transformers can be designed to implement diffusion models directly in pixel space, rather than in an autoencoder’s latent space  \citep[e.g. diffusion transformers][]{peebles2023scalable}. Moreover, operating directly in pixel space removes the need to train an additional autoencoder alongside the diffusion model, simplifying the overall training pipeline.

Our architecture is built from the JiT (Just image Transformer) architecture presented in \cite{li2026basicsletdenoisinggenerative} for pixel-space image diffusion. The input image is decomposed into patches, and each patch is projected into an embedding representation by a linear transformation. These embedding vectors, also called tokens, are summed with the patches positional embedding. This creates a sequence of tokens which are processed by several transformer blocks, returning another sequence which is finally projected back into patches with a linear layer called the predictor. 

Because we need our neural network to operate jointly on a field and on cosmological parameters, we similarly project the six parameters to cosmological embeddings of the same dimension as the patches, which we concatenate to the sequence of patch embeddings. This way, the cosmological information is just another token the transformer has to process. We use different projector and predictor layers for the cosmology and for the field patches because they do not represent the same modality, but the transformer blocks are shared. See \autoref{fig:jade} for an illustration of the architecture.

According to the denoiser parameterization described in Section~\ref{sec:jade}, the model takes as input both the noisy field $\kappa_t$ and the noisy cosmology $\theta_t$ and predicts the denoised field and cosmology. The model is also conditioned on the time $t$, providing information on the amount of noise. Further details on the architecture, such as conditioning strategy, patch sizes, transformer blocks, can be found in Appendix~\ref{sec:architecture}.

Once trained, the denoiser can be turned into a velocity field estimator with \autoref{eq:velocity-denoiser} and the unconditional distribution $p(\kappa, \theta)$ can be sampled jointly by sampling random fields $\kappa_0\sim \mathcal{N}(0, I_{d\times d})$ and parameters $\theta_0\sim \mathcal{N}(0, I_6)$, and solving \autoref{eq:ode}.

\subsection{Amortized joint posterior inference} \label{sec:amortized-posterior}

The diffusion model described above learns the unconditional joint distribution $p(\kappa, \theta)$. In practice, however, we wish to condition on an observed noisy shear field $\gamma_\text{obs}$ and sample from the posterior $p(\kappa, \theta \mid \gamma_\text{obs})$. We now describe how \jade\ can be extended to perform this conditional sampling in a fully amortized fashion, producing posterior samples for any new observation without retraining.

Sampling a conditional distribution with a diffusion model requires conditioning the velocity field on the observation, yielding $v_\varphi(\kappa_t, \theta_t, t, \gamma_\text{obs})$. With \jade, conditioning on an observed field is architecturally straightforward, because the observed field can be processed in the same way as the sampled field. The observed field $\gamma_\text{obs}$ is also split into patches, which are projected into embeddings by a dedicated linear projector, combined with positional embeddings, and concatenated to the existing sequence of target field and cosmology tokens. The full sequence is then processed jointly by the shared transformer blocks. The output tokens corresponding to the conditioning input are discarded, as they are not constrained during training and carry no physical meaning. The transformer's attention mechanism therefore provides a natural and flexible way to incorporate the conditioning information: each target token, whether a field patch or the cosmology, can attend to the observed shear at all spatial positions, allowing the model to learn which features of the data are informative for each component of the posterior.

A practical advantage of this design is that the conditioning input does not require the same patch resolution as the target field. Because the observed shear $\gamma_\text{obs}$ is contaminated by shape noise, small-scale information is suppressed compared to the noiseless convergence map that the model aims to reconstruct. We can therefore use a bigger patch size for the conditioning input, incurring negligible information loss while reducing the sequence length and the computational cost of the self-attention operations.

The conditional model is trained with the same flow matching objective as the unconditional model (\autoref{eq:loss-velocity}), with the only difference that the velocity field is now also conditioned on $\gamma_\text{obs}$. At each training step, a triplet $(\theta, \kappa, \gamma_\text{obs})$ is drawn from the forward model, the noisy interpolants $\kappa_t$ and $\theta_t$ are constructed as before, and the denoiser is trained to predict the clean samples given both the noisy state and the observed shear. Once trained, posterior samples ($\theta, \kappa) \sim p(\theta, \kappa \mid \gamma_\text{obs})$ are obtained by initializing from Gaussian noise and integrating the conditional ODE, now with the observation $\gamma_\text{obs}$ provided as a fixed input at every integration step.

Because the conditioning is handled at the architecture level rather than through an explicit likelihood, the model learns the full data-to-posterior mapping directly from simulations. This makes the approach fully amortized: a single trained model can be applied to any observed shear field without additional optimization or sampling. As for the unconditional model, no differentiable forward model is required at any stage, but only the ability to generate training triplets $(\theta, \kappa, \gamma_\text{obs})$ from the simulator.

\section{Training on LSST Y10 weak lensing maps} \label{sec:dataset}

Similarly to \cite{Lanzieri_2025, Zeghal_2025}, we simulate the convergence field of a small patch of the sky as seen at  LSST Y10 resolution from a \wcdm\, model using the \texttt{sbi\_lens}\footnote{\url{https://github.com/DifferentiableUniverseInitiative/sbi_lens}} library.  The \wcdm\, cosmological model is described by six parameters $\theta=(\Omega_c, \Omega_b, \sigma_8, h_0, n_s, w_0)$, namely the cold dark matter and baryon density parameters, the amplitude of matter fluctuations at 8 $h^{-1}$Mpc scales, the dimensionless Hubble parameter, the spectral index, and the dark energy equation-of-state parameter. The total matter density entering the convergence in Equation~\ref{eq:convergence} is then $\Omega_m = \Omega_c + \Omega_b$. Priors distribution and the fiducial values we used for this work are the same as in \cite{Lanzieri_2025}, and recalled in Table~\ref{tab:priors}.

\begin{table}

	\begin{center}
    	\begin{tabular}{lcc} 
    		\hline \hline
    		Parameter  & Prior & Fiducial value \\
    		$\Omega_c$ & $\mathcal{N}_{[0, \cdot)}$ (0.2664, 0.2) & 0.2664 \\
    		$\Omega_b$ & $\mathcal{N}$ (0.0492, 0.006) & 0.0492 \\
    		$\sigma_8$ & $\mathcal{N}$ (0.831, 0.14) & 0.831 \\
    		$h_0$ & $\mathcal{N}$ (0.6727, 0.063) & 0.6727\\
    		$n_s$ & $\mathcal{N}$ (0.9645, 0.08) & 0.9645 \\
    		$w_{0}$ &  $\mathcal{N}_{[-2,-0.3 ]}$ (-1.0, 0.9) &  -1.0 \\
    		\hline
    	\end{tabular}
        \caption{Parameters, priors and fiducial values used for our weak lensing simulations. $\mathcal{N}_\text{[a, b]}$ denotes a  normal distribution truncated on the interval $\text{[a, b]}$.}
	    \label{tab:priors}
    \end{center}
\end{table}

The Vera C. Rubin Observatory's LSST \citep{ivezic2019lsst} is a next-generation galaxy survey that will image billions of galaxies over ten years, making it one of the most powerful weak lensing datasets to date. We adopt the Y10 configuration, corresponding to the full survey depth. The distribution of the source redshift bins is parameterized by a Smail distribution 
 $n(z) \propto z^2 e^{-(z/z_0)^{\alpha}}$ \citep{smail1995deep}, with $z_0=0.11$, $\alpha=0.68$. We also consider a systematic uncertainty on the redshift with a bias parameter $\sigma_z = 0.05(1+z)$. All our fields are spatially binned into $128\times 128$ pixels, spanning a $5\times 5 \deg^2$ area, with a pixel resolution of 2.34 arcmin/pixel.

The shape noise per pixel in each tomographic bin is determined by the intrinsic ellipticity dispersion $\sigma_e = 0.26$ and the galaxy number density. Each of the 5 tomographic bins has an effective density of $\bar{n} = 27 \, \mathrm{arcmin}^{-2}$, yielding a per-pixel noise standard deviation of
\begin{equation}
    \sigma_{\mathrm{pix}} = \frac{\sigma_e}{\sqrt{\bar{n} \, A_{\mathrm{pix}}}},
\end{equation}
where $A_{\mathrm{pix}} = (5 \times 60 / 128)^2 \approx 5.49 \, \mathrm{arcmin}^2$ is the pixel area, which corresponds to averaging over approximately 148 galaxies per pixel.

\medskip

We trained \jade\, on 100,000 independent draws of $(\theta, \kappa, \gamma_\text{obs})$, from the forward model. Training is performed on a single NVIDIA GH200 GPU with 96\,GB of VRAM, using a batch size of 128 for 750 epochs, completing in approximately 48 hours. Once trained, generating a single joint posterior sample $(\theta, \kappa) \sim p(\theta, \kappa \mid \gamma_\text{obs})$ takes approximately 0.2 seconds, corresponding to the 256 ODE integration steps listed in \autoref{tab:placeholder}. Drawing 1,000 posterior samples for a given observation therefore requires roughly 3 minutes of wall-clock time. In the following section, we compare the posterior obtained with our \jade\, diffusion model and with the No-U-Turn Sampler \citep[NUTS;][]{hoffman2011nouturnsampleradaptivelysetting}. We run 10 NUTS chains in parallel, initialized at the field and cosmology ground truths, with 500 warmup steps and 3 000 sampling steps per chain, with a maximum tree depth of 6, allowing up to $2^{6}-1 = 63$ leapfrog steps per iteration, and an initial step size of $10^{-2}$. In practice the sampler saturated the maximum tree depth at essentially every iteration, averaging $62.3$ leapfrog steps
per sample. Across the 10 chains this amounts to $2.18\times10^{6}$ total gradient evaluations, or equivalently $4.36\times10^{6}$ simulation calls (two per gradient) for evaluating the gradients via automatic differentiation. This is of similar order of magnitude as the sampling benchmark in \citet{Zeghal_2025}, and one order of magnitude more than what we used to train our implicit joint posterior model \jade.

\section{Results} \label{sec:results}

We now present our results on joint posterior sampling from the conditional model $p(\theta, \kappa \mid \gamma_\text{obs})$, and verify both the quality of the reconstructed convergence fields and the recovered cosmological constraints.

\begin{figure*}
    \centering
    \includegraphics[width=1\linewidth]{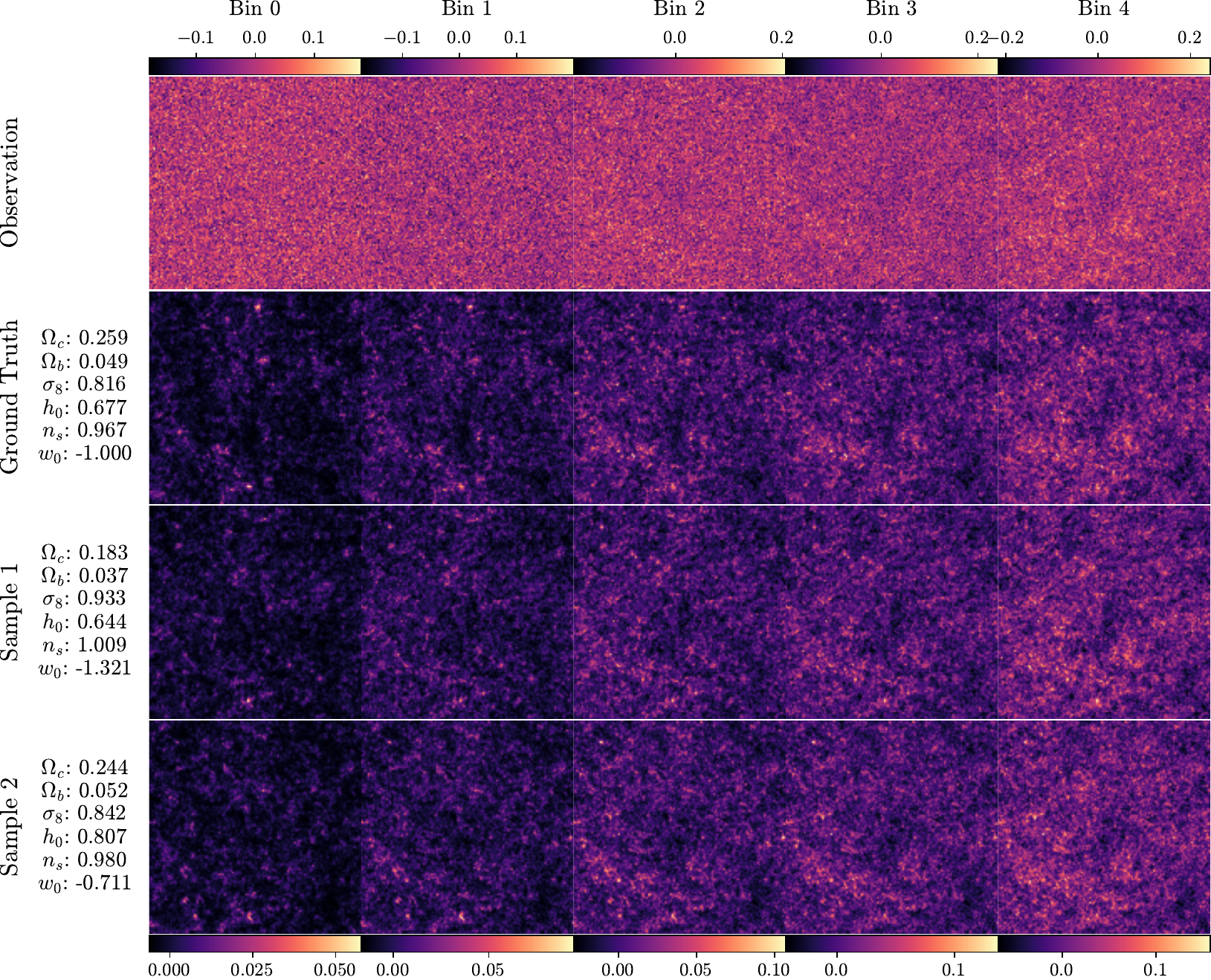}
    \caption{The first row shows the 5 bins of the observed convergence $\kappa_\text{obs}=\mathbf{F}\mathbf{P F^{-1}}\gamma_\text{obs}$, i.e. the Kaiser-Squires transformation of the observed shear field. The second row shows ground truth convergence and cosmology behind the noise. The third and fourth rows show posterior cosmology and convergence field, jointly sampled with our conditional diffusion model $(\theta, \kappa) \sim p_\varphi(\theta, \kappa \mid \gamma_\text{obs})$.}
    \label{fig:field-posterior}
\end{figure*}

We trained a conditional $\texttt{JADE}$ model, learning directly $p(\theta, \kappa \mid \gamma_\text{obs})$ . The architecture and hyperparameters are detailed in appendix \ref{sec:architecture}.

\subsection{Posterior samples validation}
\autoref{fig:field-posterior} shows joint posterior samples of the cosmological parameters and convergence maps, conditioned on a single noisy shear observation. The observed shear is visibly noisy, yet the posterior samples recover the structures of the ground-truth convergence field, while correctly reflecting the residual uncertainty on small scales.

\begin{figure}
    \centering
\includegraphics[width=\linewidth]{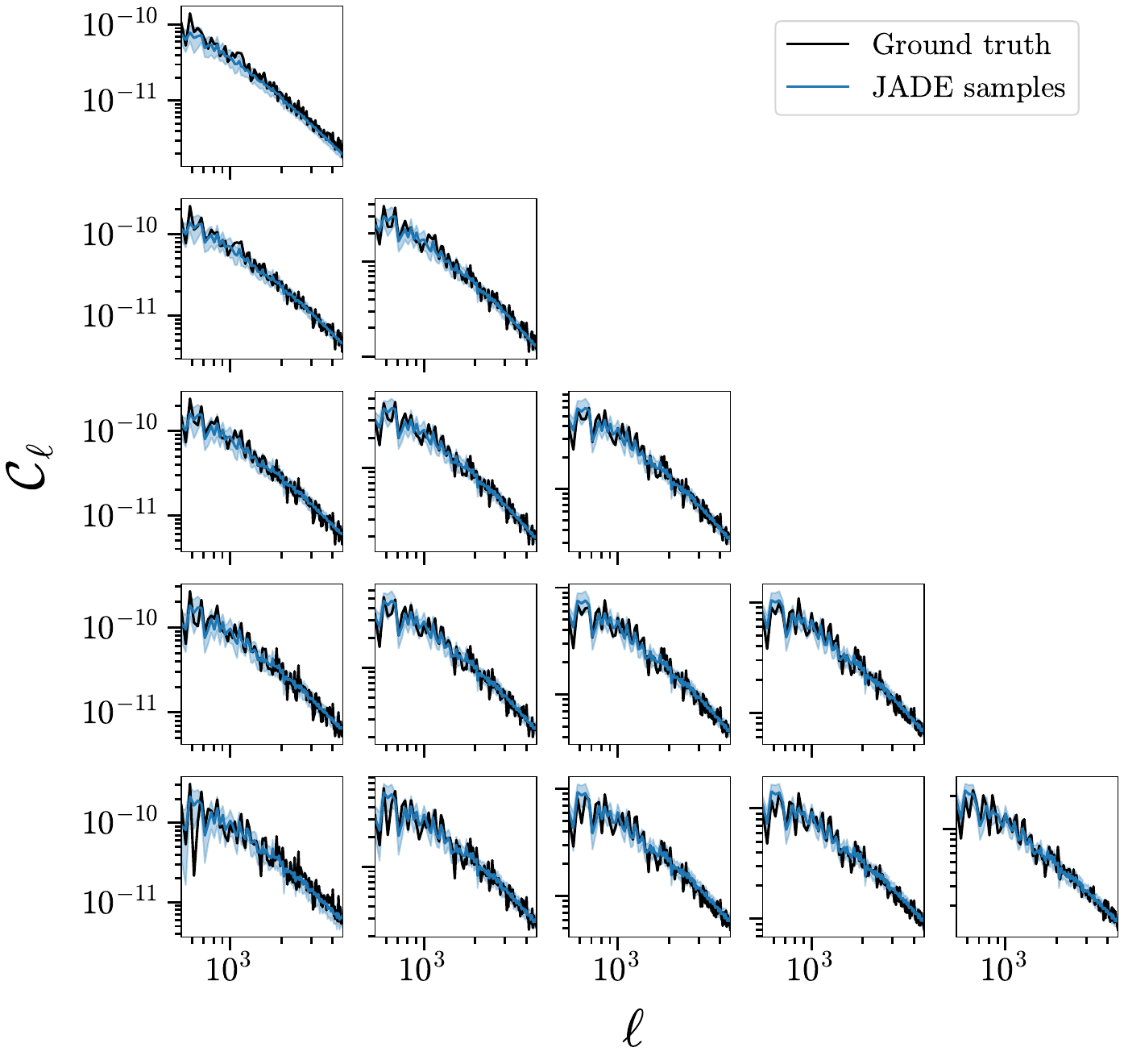}
    \caption{Auto- and cross-power spectra $\mathcal{C}_\ell$ of the convergence field for all tomographic bin combinations. For each posterior draw ($\theta^{(i)}$, $\kappa^{(i)}$), the black curves show the spectra of an independent simulation drawn at the sampled cosmology $\theta^{(i)}$, while the corresponding posterior samples $\kappa^{(i)}$ are shown in blue. The solid blue line indicates the mean over posterior samples, and the shaded region denotes the sample spread at $1\sigma$.}
    \label{fig:posterior-power-spectra}
\end{figure}

In order to verify that the reconstructed convergence fields are statistically consistent with the jointly sampled cosmology, we perform two types of checks. For each posterior sample $(\theta^{(i)}, \kappa^{(i)})$, we generate an independent simulation $\kappa_\text{sim}$ from the forward model at the sampled cosmology $\theta^{(i)}$ and compare summary statistics between $\kappa^{(i)}$ and $\kappa_\text{sim}$. \autoref{fig:posterior-power-spectra} shows that the auto- and cross- power spectra of the posterior samples match, sample by sample, those of independent simulations drawn at the corresponding $\theta^{(i)}$ across all tomographic redshift bin combinations. To quantify the overall spectral accuracy, we also compare the mean power spectrum averaged over many ground-truth convergence fields to the mean averaged over posterior samples across multiple observations. We compute the relative error between these aggregates by computing the following ratio $(\overline{\mathcal{C}}_{\ell, \text{posterior}}-\overline{\mathcal{C}}_{\ell, \text{simulation}})/\overline{\mathcal{C}}_{\ell, \text{simulation}}$. This aggregate comparison yields a relative error of $6.7\%$ averaged over all redshift bins and scales.

At the pixel level, we verify that the posterior samples are consistent with the observed data by computing the reduced chi-squared statistic
\begin{equation}
    \chi^2_\text{red} = \frac{1}{N_\text{pix}} \sum_{i,j=1}^{128} \sum_{z=1}^{5} \left( \frac{\gamma_{\text{obs},ijz} - \kappa^{(s)}_{ijz}}{\sigma_{\text{lsst},z}} \right)^2,
\end{equation}
where $N_\text{pix} = 128 \times 128 \times 5 = 81920$ and $\sigma_{\text{lsst},z}$ is the per-bin noise standard deviation. If the posterior samples faithfully reconstruct the convergence field, the residual between the observation and a posterior sample should be dominated by the injected shape noise, yielding $\chi^2_\text{red} \approx 1$. Averaged over 1000 posterior samples, we obtain $\chi^2_\text{red} = 1.0024$, consistent with the expected value of $1.0000 \pm 0.0049$, confirming that the residuals are statistically indistinguishable from the noise.

We similarly compare the one-point probability density functions (PDFs) of each posterior sample $\kappa^{(i)}$ to those of the matched simulation at $\theta^{(i)}$, finding excellent agreement across all five tomographic redshift bins (\autoref{fig:posterior-pdf}). As for the power spectrum, we also compute an aggregate PDF comparison by averaging the one-point distributions over many ground-truth fields and over posterior samples, obtaining a relative error $(\overline{p(\kappa)}_\text{posterior}-\overline{p(\kappa)}_\text{simulations})/\overline{p(\kappa)}_\text{simulations}$ of $5.2\%$ averaged across all redshift bins. Because each posterior sample is tested against a simulation at its own sampled cosmology, rather than against a fixed fiducial reference, these checks directly probe the consistency of the joint posterior, not just the marginal field statistics. The recovered PDFs additionally exhibit the expected log-normal shape, confirming that the model has learned the non-Gaussian structure imposed by the log-normal forward model. Further statistical comparisons are presented in Appendix~\ref{sec:statistics-comparison}.

\subsection{Posterior coverage validation}

\begin{figure}
    \centering
    \includegraphics[width=1.\linewidth]{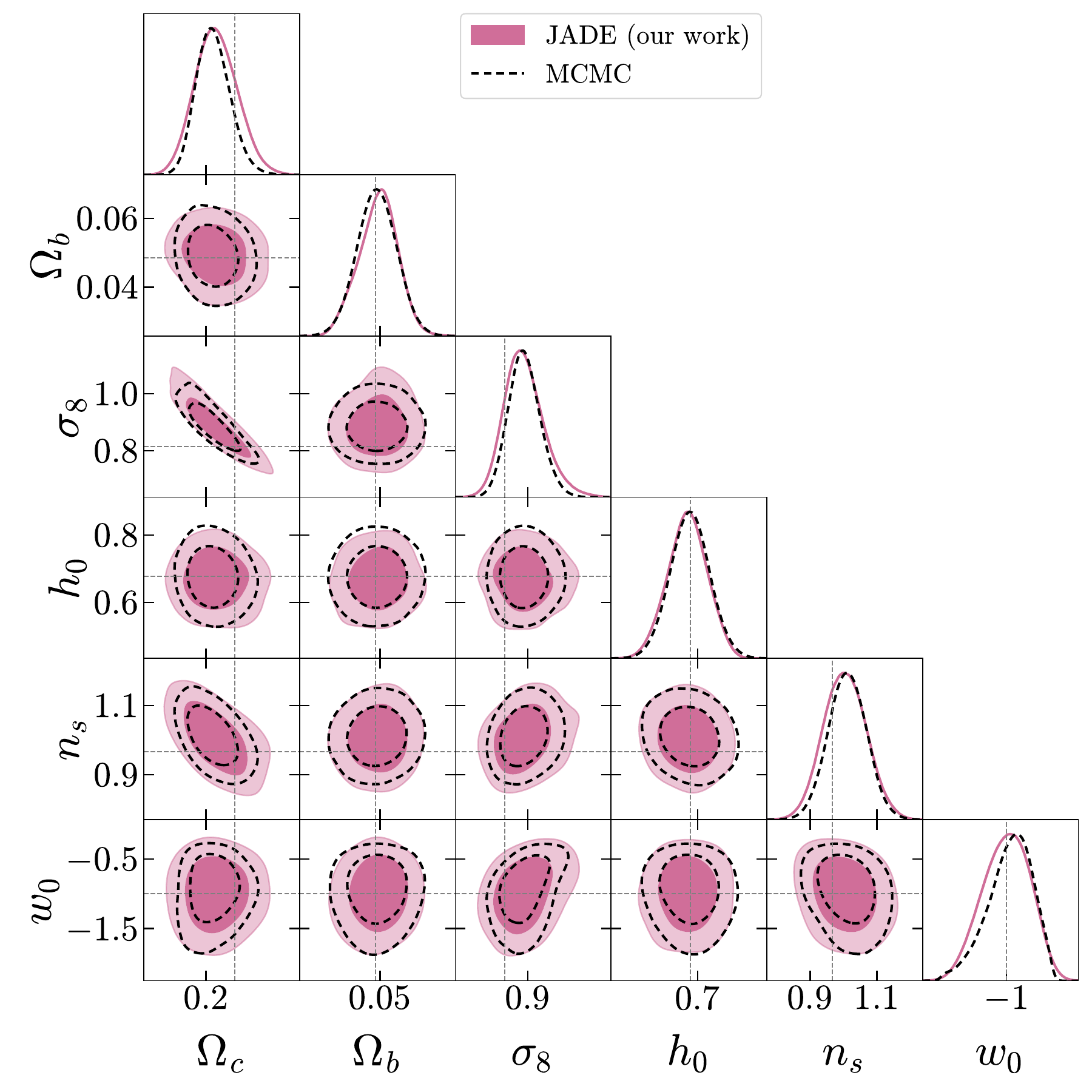} \\
    \caption{1D and 2D marginal posterior $p(\theta \mid \gamma_\text{obs})$ contour plot, comparing our amortized diffusion sampling and classical MCMC contours for a single observation generated at fiducial Planck cosmology.
    }
    \label{fig:marginal-posterior}
\end{figure}

\autoref{fig:marginal-posterior} shows the 1D and 2D marginal posterior $p(\theta\mid \gamma_\text{obs})$, obtained by marginalizing the field from our joint posterior samples. For comparison, we run MCMC using NUTS, which requires gradients of the log-normal forward model provided by the differentiability of the \texttt{sbi\_lens} library. The diffusion-based contours are in excellent agreement with the MCMC result, recovering consistent credible regions and degeneracy directions across all parameter pairs.  We note that while NUTS requires a differentiable forward model and must be run independently for each observation, our diffusion model is trained only from forward simulations and is amortized over observations. We further discuss the practical implications of these differences in Section \ref{sec:discussion}. 

\begin{figure}
    \centering
    \includegraphics[width=.9\linewidth]{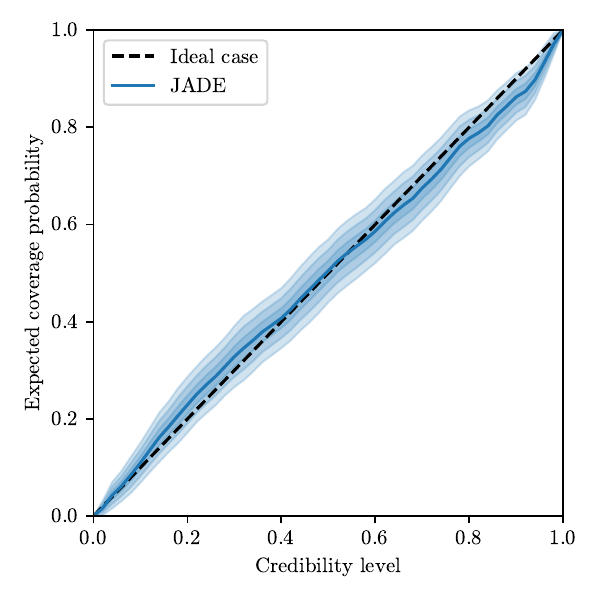}
    \caption{TARP coverage test of the marginal posterior $p(\theta \mid \gamma_\text{obs})$ inferred with the amortized model, showing mean and error bars with $\sigma \in [1,2,3]$ computed with bootstrap.}
    \label{fig:tarp-theta-amortized}
\end{figure}

We assess the calibration of the estimated posteriors using MIRA (Mass In Random Areas), a score for evaluating the accuracy of conditional distributions \citep{sharief2026mirascoreconditionaldistribution}, together with the Tests of Accuracy with Random Points \citep[TARP;][]{lemos2023sampling} diagnostic. Both the MIRA and TARP diagnostics are computed using $500$ observations with $500$ posterior samples drawn per observation. MIRA is a sample-based statistical test that verifies whether a conditional distribution assigns the correct density to any region of parameter space, by comparison with simulated parameters $\theta, \kappa, \gamma_\text{obs} \sim p(\theta, \kappa, \gamma_\text{obs})$. A perfectly calibrated posterior yields a MIRA score of $2/3$, while the worst-case score is $1/2$. Applying MIRA to our learned joint posterior $p(\theta, \kappa \mid \gamma_\text{obs})$, we obtain a score of $0.635 \pm 0.017$, indicating good calibration. This value suggests that the posterior is mildly overconfident, but it represents a strong result for a high-dimensional distribution. We further compute the MIRA score on the marginal cosmological posterior $p(\theta \mid \gamma_\text{obs})$ and obtain $0.659 \pm 0.0139$, which lies within MIRA's theoretical error bars of $1/\sqrt{18L}$, where $L$ is the number of conditioning observations $\gamma_\text{obs}$ used to estimate the score, and reflects even better calibration. TARP complements this analysis by evaluating whether the credible regions of the inferred posterior achieve the correct frequentist coverage across many test observations. \autoref{fig:tarp-theta-amortized} shows the resulting coverage plot, with the mean and $1\sigma$, $2\sigma$, and $3\sigma$ error bars computed via bootstrap resampling. The coverage curves are consistent with the diagonal, indicating that the marginal cosmological posterior is well calibrated across the parameter space.

\section{Discussion and conclusion} \label{sec:discussion}

In this work, we introduced \jade, a joint architecture for the simultaneous inference of convergence fields and cosmological parameters from weak lensing observations. By learning the joint posterior $p(\theta, \kappa \mid \gamma_\text{obs})$ with a single conditional diffusion model, our method unifies mass map reconstruction and cosmological parameter estimation within a coherent Bayesian framework, without requiring a differentiable forward model or MCMC sampling at inference time.

We validated our approach on a simulated LSST Y10-like weak lensing setting using log-normal convergence fields in a \wcdm\, model. The reconstructed convergence maps are statistically consistent with the jointly inferred cosmology, as verified through power spectra and one-point PDF comparisons. The marginal cosmological posterior is in excellent agreement with NUTS-based MCMC chains and is well calibrated as assessed by the TARP and MIRA diagnostics. These results demonstrate that a single diffusion model can jointly recover both the field and the cosmology at a quality comparable to dedicated methods for each task individually.

Our method offers several practical advantages over explicit inference approaches, which as discussed in Section~\ref{sec:field-level}, require a differentiable forward model and MCMC sampling.
Training requires only forward simulations, not a differentiable simulator, making it in principle applicable to arbitrarily complex forward models including full N-body and hydrodynamical codes for which analytical gradients are unavailable. 
The model is amortized: once trained, posterior samples for any new observation are obtained by integrating the learned ODE at a cost of approximately 0.2 seconds per sample, and training needs one order of magnitude fewer simulations than required for sampling.

Several directions remain for future improvement. Operating the diffusion model directly in pixel space, as we do here, becomes computationally challenging as the field resolution increases, since the transformer sequence length grows with the number of patches. This can be mitigated by adopting a latent diffusion approach \citep{rombach2022high}, in which an autoencoder first compresses the field into a lower-dimensional representation on which the diffusion model operates. Such a strategy would significantly reduce the cost of both training and sampling, enabling application to higher-resolution fields at the expense of an additional autoencoder training stage.

Our current demonstration relies on log-normal simulations, which are fast enough to generate the large training datasets required by the diffusion model. Scaling to more realistic but computationally expensive simulators, such as full N-body codes, introduces additional challenges. The training set must densely cover the cosmological parameter space to ensure the learned posterior generalizes across cosmologies, and the cost of generating sufficient simulations may become prohibitive. Furthermore, as an amortized method, \jade, must be retrained whenever the data model changes, for instance when updating the survey mask, noise model, or systematic effects, unlike explicit inference methods which can adapt at inference time. Our setting also assumes a flat-sky approximation on a small patch; extending to curved-sky geometries and larger survey footprints will require architectural adaptations. More broadly, implicit inference methods rely on the fidelity of the training simulations: any mismatch between the simulated and real data distributions will propagate directly into the inferred posterior, making careful validation of the forward model essential before application to observational data.

Although our method does not require a differentiable simulator, it can benefit from one when available. As proposed by \cite{zeghal2022neural, Zeghal_2025}, gradient information from a differentiable forward model can be incorporated during training to improve sample efficiency, reducing the number of simulations needed to train the diffusion model. This is particularly relevant when training on expensive N-body simulations, where each sample carries a significant computational cost. Notably, this gradient-informed training strategy is not straightforward to apply to standard simulation-based inference methods for cosmology, which typically learn an amortized marginal posterior $p(\theta \mid \gamma_\text{obs})$: the simulator gradients are defined on the joint space of fields and parameters, and there is no direct way to compare them against the output of a model that has marginalized over the fields. Because \jade\, models the full joint posterior $p(\theta, \kappa \mid \gamma_\text{obs})$, the learned denoiser operates in the same space as the simulator gradients, making it possible to incorporate this information directly into the training objective. We leave the exploration of this direction to future work.

Finally, we note that the framework presented here is not specific to convergence field reconstruction. The same architecture and training procedure can be applied to any field-level inference problem where both a physical field and global parameters must be jointly recovered from observations. A natural extension is the reconstruction of initial conditions of the Universe jointly with cosmological parameters, which we will explore in future work. More broadly, \jade\, can be applied to other cosmological probes and combined with multi-probe analyses, where the joint inference of shared fields and parameters across different observables could further tighten cosmological constraints. 

\section*{Acknowledgements}
\addcontentsline{toc}{section}{Acknowledgements}

All authors gratefully acknowledge the support of the NSF-Simons AI-Institute for the Sky (SkAI) via grants NSF AST-2421845 and Simons Foundation MPS-AI-00010513. This research used the DeltaAI advanced computing and data resource, which is supported by the National Science Foundation (award OAC 2320345) and the State of Illinois. DeltaAI is a joint effort of the University of Illinois Urbana-Champaign and its National Center for Supercomputing Applications.

\medskip

The code written to produce the results presented in this paper uses the following software: JAX \citep{jax2018github}, Flax \citep{flax2020github}, Numpy \citep{harris2020array}, Matplotlib \citep{Hunter:2007}, Lenstools \citep{Petri_2016}, \sbilens\ \citep{Lanzieri_2025}.


\bibliographystyle{mnras}
\bibliography{references} 


\appendix

\section{Architecture details}\label{sec:architecture}

We describe here the details of the \jade\, architecture, its training recipe, and the sampling procedure. The full set of hyper-parameters is reported in \autoref{tab:placeholder}. Our design closely follows the Just image Transformer \citep[JiT;][]{li2026basicsletdenoisinggenerative}, extended to operate jointly on a convergence field and on cosmological parameters, and optionally on an observed shear field used as conditioning.

\begin{table}
    \centering
    \begin{tabular}{lc}
    \hline
    \hline
        \textbf{Architecture} & \\
        image size & $128\times 128$ \\
        channels (tomographic bins) & 5 \\
        patch size $p$ & 8 \\
        conditioning patch size $p_\text{cond}$ & 8 \\
        \# cosmology tokens $N_\theta$ & 16 \\
        depth $L$ & 12 \\
        hidden dim $d$ & 768 \\
        attention heads $h$ & 12 \\
        bottleneck dim $b$ & 128 \\
        normalization & RMSNorm \\
        feed-forward & SwiGLU \\
        positional embedding (field/cond) & RoPE \\
        attention stabilization & qk-norm \\
        time conditioning & adaLN-Zero \\
        \hline
        \textbf{Training} & \\
        loss type & $v$-loss ($x$-pred.) \\
        time distribution & logit-normal \\
        $(\mu, \sigma)$ & $(-0.8,\, 0.8)$ \\
        $(1-t)$ clipping $t_\text{eps}$ & 0.05 \\
        optimizer & AdamW \\
        $(\beta_1, \beta_2)$ & $(0.9,\, 0.95)$ \\
        weight decay & 0 \\
        schedule & warmup + cosine decay \\
        initial learning rate & $5\times 10^{-3}$ \\
        final learning rate & $1\times 10^{-4}$ \\
        data augmentation & flips + $90^{\circ}$ rotations \\
        noise resampling & on the fly \\
        gradient norm clipping & 1.0 \\
        EMA decay & 0.999 \\
        batch size & 128 \\
        epochs & 750 \\
        precision & bfloat16 \\
        \hline
        \textbf{Sampling} & \\
        solver & Heun \\
        steps & 256 \\
        time grid & linear in $[0, 1]$ \\
        \hline
    \end{tabular}
    \caption{Hyper-parameters used to set up the architecture, train, and sample from the conditional \jade\ model.}
    \label{tab:placeholder}
\end{table}

\subsection{Patch embeddings and token sequence}

The target convergence field $\kappa \in \mathbb{R}^{H\times W\times C}$, with $H=W=128$ and $C=5$ tomographic bins, is divided into non-overlapping patches of size $p\times p$ with $p=8$, yielding a sequence of $(H/p)\times(W/p) = 16\times 16 = 256$ field tokens. Each patch is a $p\times p\times C = 320$-dimensional vector, which is projected into the $d=768$-dimensional hidden space of the transformer by a bottleneck linear embedding \citep{li2026basicsletdenoisinggenerative}. Following their prescription, we replace the single linear embedding of the standard ViT \citep{dosovitskiy2021imageworth16x16words} by two successive linear layers with an intermediate bottleneck dimension $b=128$, acting as a low-rank reparameterization of the patch embedding. The symmetric predictor that projects output tokens back to $p\times p\times C$ patches also uses a bottleneck pair of linear layers. Fixed sinusoidal 2D positional embeddings are added to the field patch embeddings.

The six cosmological parameters $\theta\in\mathbb{R}^{6}$ are projected to a sequence of $N_\theta = 16$ tokens in the same hidden space by a dedicated linear projector, to which we add learnable positional embeddings. Using multiple cosmology tokens rather than a single one follows the in-context class conditioning strategy of \cite{li2026basicsletdenoisinggenerative} (repeated from MAR, \citealt{li2024autoregressive}), and gives the transformer a larger capacity to propagate cosmological information through the sequence. At the output, the corresponding $N_\theta$ tokens are averaged and passed through a separate linear head to predict the denoised cosmology $\theta$.

When a noisy shear observation $\gamma_\text{obs}\in\mathbb{R}^{H\times W\times C}$ is provided as conditioning, it is tokenized with an independent bottleneck patch embedding and its own learnable 2D positional embedding. This produces an additional $16\times 16 = 256$ conditioning tokens. The full input sequence to the transformer is therefore
\begin{equation}
    [\underbrace{e^\theta_1, \dots, e^\theta_{16}}_{\text{cosmology}}, \; \underbrace{e^{\gamma}_{1}, \dots, e^{\gamma}_{64}}_{\text{conditioning}}, \; \underbrace{e^{\kappa}_{1}, \dots, e^{\kappa}_{256}}_{\text{field}}],
\end{equation}
for a total length of $528$ tokens. The output tokens associated with the conditioning are discarded, as they are not constrained by the loss.

\subsection{Transformer blocks}

The token sequence is processed by $L=12$ transformer blocks with hidden dimension $d=768$ and $h=12$ attention heads, following the recipe of \cite{li2026basicsletdenoisinggenerative}. Each block consists of multi-head self-attention followed by a feed-forward network, with pre-normalization. We adopt the following modern components: RMSNorm \citep{zhang2019root} in place of LayerNorm, SwiGLU \citep{shazeer2020glu} as the feed-forward non-linearity, rotary positional embeddings \citep[RoPE;][]{su2024roformer} applied to the field and conditioning tokens (the cosmology tokens are excluded from RoPE and act as a register-like prefix), and query-key normalization \citep{henry2020query} for attention stability. Formally, a single attention head computes
\begin{equation}
    \text{Attn}(Q, K, V) = \text{softmax}\!\left(\frac{\mathcal{R}(Q)\,\mathcal{R}(K)^\top}{\sqrt{d_h}}\right) V,
\end{equation}
where $Q, K, V$ are the query, key, and value projections of the normalized input, $d_h = d/h = 64$ is the per-head dimension, and $\mathcal{R}$ denotes the RoPE rotation applied to field and conditioning tokens. Queries and keys are additionally RMSNorm-normalized before the attention product.

\subsection{Time conditioning}

The diffusion time $t\in[0,1]$ is encoded with a sinusoidal frequency embedding followed by a two-layer MLP, producing a conditioning vector $c(t)\in\mathbb{R}^{d}$. This vector modulates every transformer block and the field predictor head via adaLN-Zero \citep{peebles2023scalable}. Inside each block, the scale, shift, and residual gate parameters of the attention and MLP sublayers are predicted as affine functions of $c(t)$. The adaLN projections are initialized to zero so that each block starts as an identity mapping at the beginning of training, which has been shown to stabilize the optimization of diffusion transformers \citep{peebles2023scalable}.

\begin{figure*}
    \centering
    \includegraphics[width=\linewidth]{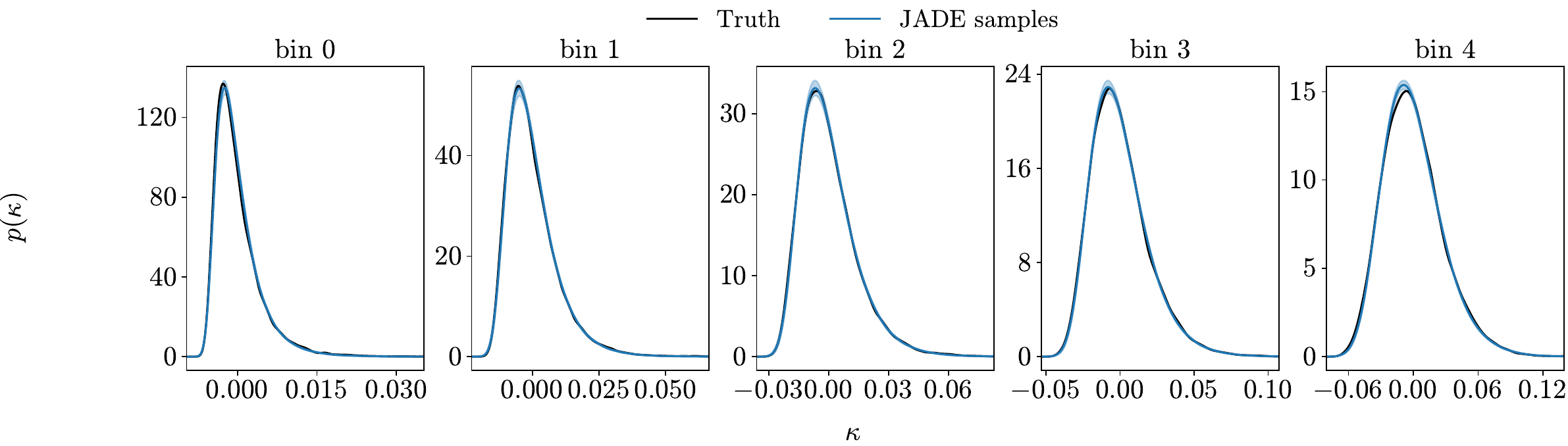}
    \caption{One-point probability density functions of the convergence field for each of the five tomographic bins. For each posterior draw $(\theta^{(i)}, \kappa^{(i)})$, the posterior sample PDF (blue) is compared to the PDF of an independent simulation generated at the sampled cosmology $\theta^{(i)}$ (black). The posterior samples closely reproduce the expected log-normal shape across all bins, confirming that the model captures the non-Gaussian pixel statistics of the convergence field beyond the two-point information validated by the power spectrum.}
    \label{fig:posterior-pdf}
\end{figure*}

\subsection{Initialization}

Linear layers are initialized with Xavier uniform initialization \citep{glorot2010understanding}. Following \cite{peebles2023scalable} and \cite{li2026basicsletdenoisinggenerative}, the adaLN modulation layers and the final predictor layers are zero-initialized, so that the network initially predicts the clean sample equal to the noisy input (and the corresponding velocity is the identity interpolation), providing a stable starting point for flow matching training.

\subsection{Time distribution and loss}

Following \cite{esser2024scaling} and \cite{li2026basicsletdenoisinggenerative}, during training we sample the diffusion time $t$ from a logit-normal distribution: $s\sim\mathcal{N}(\mu,\sigma^2)$ and $t = \text{sigmoid}(s)$, with $\mu=-0.8$ and $\sigma=0.8$. This biases training towards mid-to-high noise levels, which we empirically found beneficial in our setting.

Following the trick advocated by \cite{li2026basicsletdenoisinggenerative}, we optimize the flow matching $v$-loss of \autoref{eq:loss-velocity}, but parameterize the neural network so that it predicts the clean sample $x_1 = [\theta_1, \kappa_1]$ rather than the velocity. The velocity $v_\varphi$ entering the loss is reconstructed from the denoiser output $d_\varphi$ through \autoref{eq:velocity-denoiser}, so that the training objective reads
\begin{equation}\label{eq:jade-loss}
    \mathcal{L}(\varphi) = \mathbb{E} \left[ \left\| \frac{d_\varphi(x_t, t, \gamma_\text{obs}) - x_t}{\max(1-t,\, t_\text{eps})} - (x_1 - x_0) \right\|^2 \right],
\end{equation}
where the expectation is taken over $p(t)\,p(x_1)\,p(x_0)$ with $x_t = t x_1 + (1-t) x_0$, and $t_\text{eps}=0.05$ clips the $(1-t)$ denominator as $t\to 1$ \citep{li2026basicsletdenoisinggenerative}. We emphasize that the field and cosmology components of the denoiser enter this single, unweighted mean-squared-error objective on the same footing. Introducing a relative weight between the field and cosmology components of the velocity residual in \autoref{eq:jade-loss} would no longer target the true joint velocity field, and would therefore bias the recovered joint posterior $p(\kappa,\theta\mid\gamma_\text{obs})$.

\subsection{Data augmentation and on-the-fly noise}

Because the convergence field is statistically isotropic on the small patches considered here, we apply random flips and $90^{\circ}$ rotations to each clean convergence map at every training step. This cheap augmentation effectively multiplies the number of distinct fields seen by the network, reducing the risk of the transformer memorizing individual training simulations.

The conditioning shear $\gamma_\text{obs}$ is not stored alongside the dataset but is re-generated at every training iteration by applying the Kaiser-Squires operator to the convergence and adding a fresh realization of LSST-Y10 shape noise (see Section~\ref{sec:log-normal}). As a result, the network sees a different noise realization of the same underlying field at each epoch, which acts as an additional regularizer and forces the learned posterior to marginalize explicitly over the noise distribution rather than over a fixed empirical sample.

\subsection{Optimization}

\jade\, is trained with Adam \citep{kingma2014adam} using $(\beta_1, \beta_2) = (0.9, 0.95)$, no weight decay, and gradient norm clipping at $1.0$. The learning rate follows a warmup-then-cosine-decay schedule. It is linearly warmed up from $5\times 10^{-3}$ to a peak value and then decayed with a cosine profile down to $1\times 10^{-4}$ by the end of training. We use bfloat16 mixed precision on a single NVIDIA GH200 96\,GB GPU with a batch size of 128, training for 750 epochs on 100,000 simulated pairs $(\theta, \kappa)$, with $\gamma_\text{obs}$ resampled on the fly as described above. We maintain an exponential moving average of the weights with decay $0.999$, which is used for all reported results.

\subsection{Sampling}

Posterior samples are obtained by integrating the ODE in \autoref{eq:ode} with the velocity field reconstructed from the denoiser through \autoref{eq:velocity-denoiser}. We use Heun's second-order solver \citep{karras2022elucidating} with $256$ steps on a linear time grid in $[0, 1]$, starting from $\kappa_0\sim\mathcal{N}(0, I)$ and $\theta_0\sim\mathcal{N}(0, I_6)$, and with $\gamma_\text{obs}$ provided as a fixed conditioning input at every step.

\section{Posterior samples statistics check}\label{sec:statistics-comparison}

To verify that the convergence fields produced by our conditional diffusion model are statistically consistent with the jointly sampled cosmology, we perform checks of summary statistics checks. For each posterior draw $(\theta^{(i)}, \kappa^{(i)}) \sim p_\varphi(\theta, \kappa \mid \gamma_\text{obs})$, we draw an independent simulation $\kappa^{(i)}_\text{sim}$ from the forward model at the sampled cosmology $\theta^{(i)}$ and compare summary statistics between $\kappa^{(i)}$ and $\kappa^{(i)}_\text{sim}$. Because the reference simulation is drawn at $\theta^{(i)}$ rather than at a fixed fiducial cosmology, this test directly probes the consistency of the joint posterior. For the log-normal forward model considered here, the field statistics are fully specified by the power spectrum and the shift parameter $\lambda(z,\theta)$, so the auto- and cross-power spectra (\autoref{fig:posterior-power-spectra}) together with the one-point PDFs (\autoref{fig:posterior-pdf}) are statistically sufficient to validate our field-level posterior samples. Extending these checks with summaries more sensitive to scale-coupled non-Gaussianity, such as the scattering transform, will become informative when moving to N-body simulators.


\bsp	
\label{lastpage}
\end{document}